\documentclass[conference]{IEEEtran}

\IEEEoverridecommandlockouts
\usepackage{amsmath,amssymb,amsfonts}
\usepackage{algorithmic}
\usepackage{graphicx}
\usepackage{textcomp}
\usepackage{xcolor}
\usepackage{subcaption}

\usepackage{cleveref}
\usepackage{listings}

\usepackage{cite}

\newcommand{\ignore}[1]{}

\newif\ifanonymous
\anonymousfalse

\def\BibTeX{{\rm B\kern-.05em{\sc i\kern-.025em b}\kern-.08em
    T\kern-.1667em\lower.7ex\hbox{E}\kern-.125emX}}
\begin{document}

\title{Communication-Aware Diffusion Load Balancing for Persistently Interacting Objects
\\
    \ifanonymous
    \else
        \thanks{
            This material is based upon work supported by the U.S. Department of Energy, Office of Science, Office of Advanced Scientific Computing Research, Department of Energy Computational Science Graduate Fellowship under Award Number DE-SC0025528.

            This report was prepared as an account of work sponsored by an agency of the United States Government. Neither the United States Government nor any agency thereof, nor any of their employees, makes any warranty, express or implied, or assumes any legal liability or responsibility for the accuracy, completeness, or usefulness of any information, apparatus, product, or process disclosed, or represents that its use would not infringe privately owned rights. Reference herein to any specific commercial product, process, or service by trade name, trademark, manufacturer, or otherwise does not necessarily constitute or imply its endorsement, recommendation, or favoring by the United States Government or any agency thereof. The views and opinions of authors expressed herein do not necessarily state or reflect those of the United States Government or any agency thereof.

        }
    \fi

}

\ifanonymous
\else
    \author{\IEEEauthorblockN{Maya Taylor, Kavitha Chandrasekar, Laxmikant V. Kale}
        \IEEEauthorblockA{\textit{Department of Computer Science} \\
            \textit{University of Illinois at Urbana-Champaign}\\
            Champaign, IL, USA \\
            \{mayat4, kchndrs2, kale\}@illinois.edu}
    }
\fi
\maketitle

\begin{abstract}
    Parallel applications with irregular and time-varying workloads often suffer from load imbalance. Dynamic load balancing techniques address this challenge by redistributing work during execution. We present a new type of distributed diffusion-based load balancing targeted at communication-intensive applications with persistently communicating objects. Leveraging the application's communication graph, our strategy reduces across-node communication while simultaneously distributing load effectively. We also propose an algorithmic variant for cases where the communication patterns are not readily available. We explore optimizations to our algorithm, and comparisons with other related load balancing strategies in simulation and on a Particle-in-Cell benchmark on up to 8 nodes of Perlmutter at NERSC.
\end{abstract}


\section{Introduction}
\label{sec:intro}

Irregular parallelism characterizes many scientific workloads, from particle simulations and adaptive mesh refinement to graph analytics. These applications exhibit time-varying load imbalance and communication patterns, which both present performance challenges. Particularly at strong-scaling limits, where the work per processor is small and communication costs dominate, even modest load imbalance or disruption of communication locality can significantly degrade performance.

To address these challenges, many systems rely on runtime load balancing. We focus particularly on runtime systems with a chunk-based or 'over-decomposed' execution model, in which computation is divided into more work-and-data units, or \textit{objects}, than processors. These objects can be migrated dynamically, enabling the system to redistribute work as the application evolves.

Within this model, we present a new distributed load balancing algorithm inspired by diffusion-based techniques. Our approach targets the over-decomposition model of the Charm++ runtime system \cite{charm++}, but it is broadly applicable to any chunk-based programming framework that supports runtime migration, including those used in adaptive mesh refinement codes and systems such as the DARMA/vt runtime \cite{darma}. Our algorithm balances load by migrating, or \textit{diffusing}, work units between nodes that already communicate. By aligning load balancing decisions with existing communication patterns, the method preserves communication locality while achieving balanced workload distribution.

\begin{figure}[]
    \centering
    \includegraphics[
        width=0.22\textwidth,
        clip,
        trim=1.4in 2.5in 1.35in 2.5in
    ]{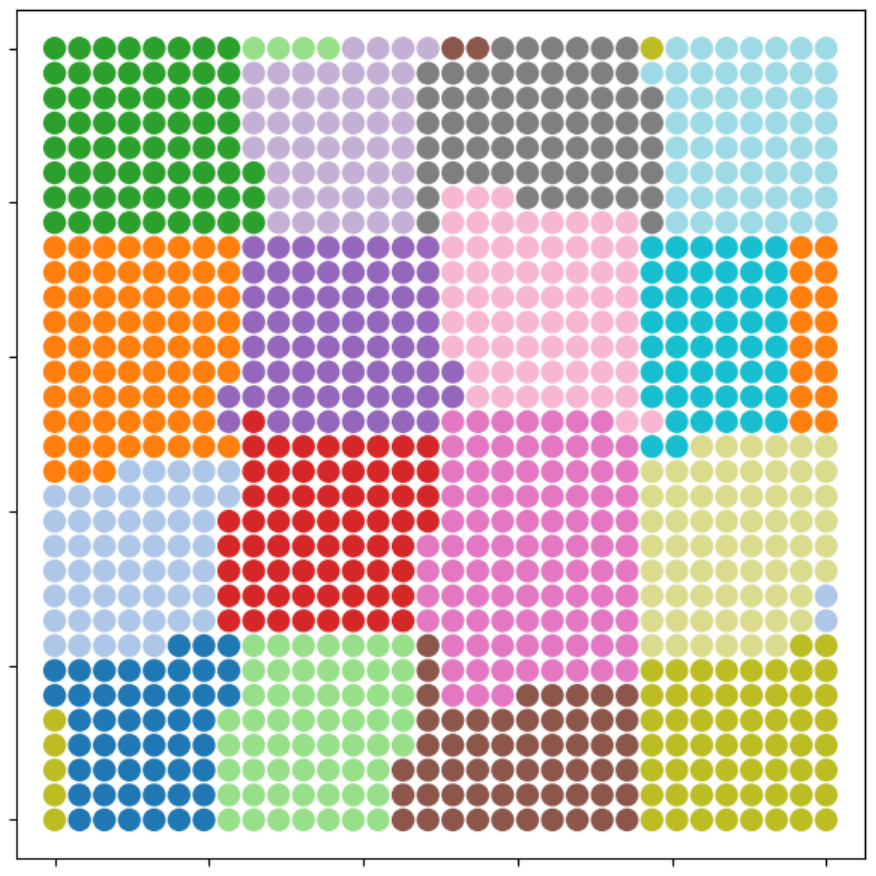}
    \includegraphics[
        width=0.22\textwidth,
        clip,
        trim=1.4in 2.5in 1.35in 2.5in
    ]{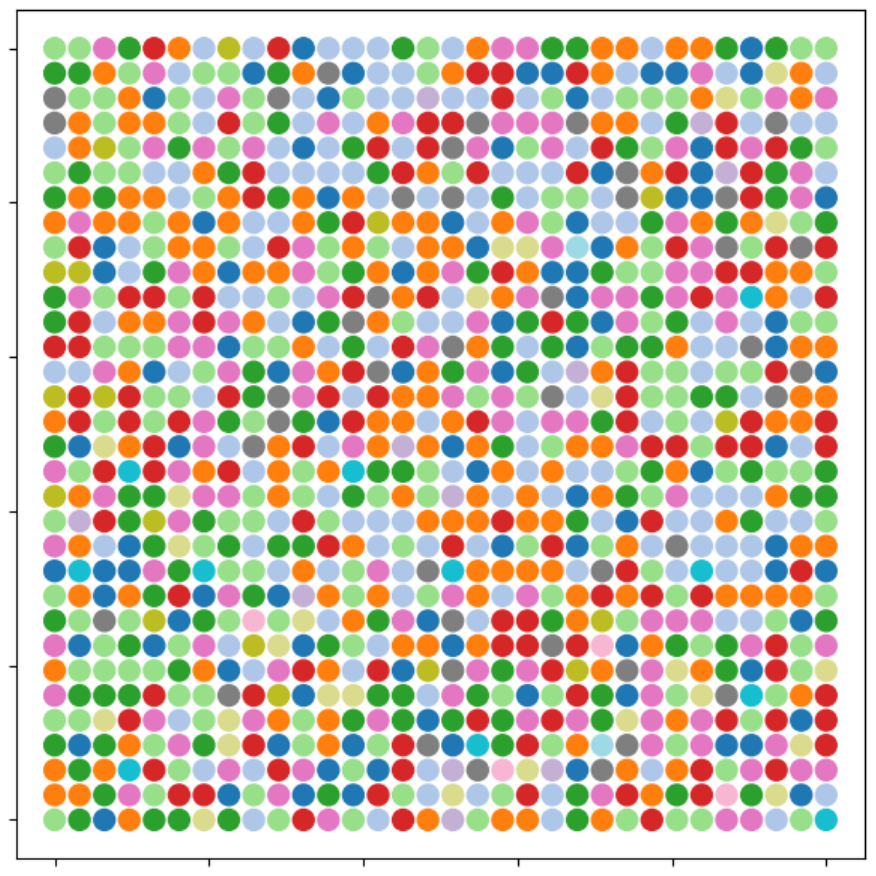}
    \caption{Load visualizations of a synthetic 2D stencil application. Circles denote migratable objects, and colors denote the owning processor. These figures were generated using our load balancing simulation infrastructure for diffusion (left) and greedy-refine strategies (right). }
    \label{fig:intro-viz}
\end{figure}

To provide intuition for our diffusive scheme, we use an example 2D stencil benchmark and corresponding visualization. Consider an application in which a 2D grid of objects, which we will refer to by their Charm++ terminology as \textit{chares}, participate in an iterative 5-point stencil algorithm. Each 2D grid point updates its value based on its own state and the states of its four immediate neighbors (north, south, east, and west). For the moment, we assume that each grid point is allocated to its own chare. While we will use this stencil framework as a simple tool to understand and simulate load balancing, stencils are also a very common computational pattern across scientific applications, including finite-difference solvers and wave propagation models.

\Cref{fig:intro-viz} presents two visualizations of a stencil application to give intuition for communication-locality. In the left figure, cells and their neighbors are frequently on the same processor, producing contiguous blocks of color. This represents good communication locality, as most of the neighbor-to-neighbor communication is local. The right figure represents the opposite, in which objects are dispersed randomly and communication locality is entirely disrupted.

Our proposed load balancing scheme aims to prevent this disruption by maintaining communication locality throughout the load balancing process.
We present two variations of our algorithm, evaluate it both in simulation and on a Particle-in-Cell (PIC) benchmark, and identify strengths of our approach and opportunity for further work.


\section{Related Work and Problem Definition}
\label{sec:related}

Several distinct categories of load balancing problems arise in parallel computing for dynamic, irregular parallelism. Two commonly studied formulations are: (1) a simple work-list formulation in which each processor holds a set of independent tasks, each with an associated computational load \cite{cybenko_dynamic_1989,menon_distributed_2013}, and (2) divide-and-conquer computations in which tasks are short-lived but continuously generated during execution \cite{lin-keller,saletore}. 
Both have been extensively studied in the literature, including many approaches inspired by diffusion processes \cite{elsasser_diffusion_2002, canright_chemotaxis-inspired_2006, deng_heat_2010}.

However, these formulations generally assume that work items do not communicate with each other (except for limited parent–child interactions in divide-and-conquer settings). 
In contrast, we focus on a category in which a set of long-lived objects persist throughout execution and repeatedly communicate with a small subset of other objects. 

{\bf Problem Definition:} We are given a set of interacting objects, each with a current processor assignment and an associated computational load, and a sparse graph of weighted communication edges representing interactions between objects. The objective is to compute a new mapping of objects to processors. The mapping is evaluated using several cost metrics: (1) load imbalance, measured as the ratio of maximum to average processor load (where load is approximated by the runtime spent within an object), (2) communication cost, measured as the ratio of inter-node communication to intra-node communication (in bytes), (3) migration cost, representing the overhead of moving objects to new processors, and (4) the cost of computing the mapping itself. 



In addressing these specific constraints, our work differs substantially from work addressing the previously described categories, which includes most existing diffusion-based approaches. The primary exception is recent work that incorporates communication-awareness into distributed load balancing, addressing relevant constraints but without diffusion \cite{lifflander_accelerating_2025, lifflander_communication-_2024}. 
Interestingly, traditional graph partitioning tools such as METIS and ParMETIS can also be used for load balancing in this context, particularly when an explicit communication graph is available \cite{metis, parmetis}. ParMETIS in particular supports a “re-partitioning” algorithm, which can partition the object communication graph iteratively throughout runtime with minimal migrations, while preserving communication locality. Lieber et al. \cite{lieber_potential_2016} provide a comprehensive overview of multiple diffusive strategies and compare them against graph-partitioning strategies, concluding that diffusion can provide substantial benefit, especially with regard to minimizing migrations, communication, and re-balancing strategy time.

We build on this array of previous work to develop a load balancing strategy for communication-intensive applications in an over-decomposed ecosystem, using broad principles of diffusion. We optimize for load balance across nodes while keeping migrations minimal and preserving communication locality.

\section{Three-stage Communication-Aware Diffusion}
\label{sec:algorithm}

Our algorithm involves (1) constructing a node neighbor graph based on an application's communication patterns, (2) computing the ideal load redistribution without considering object granularity, and (3) selecting specific objects for migration. Each of these phases is discussed in detail below.

Critically, we note that any incremental refinement strategy like ours assumes that the initial object distribution has good communication locality. If this is not guaranteed, an initial graph-partitioning load balancing pass (e.g. via METIS) can be performed to induce this locality. Additionally, adaptive runtime strategies like ours assume some persistence of load and communication patterns across load balancing iterations.

\subsection{Neighbor Selection}
Our algorithm differs from previous work in that the initial graph is constructed based on communication patterns, as opposed to network topology \cite{lieber_potential_2016}. A neighbor selection phase implicitly constructs the node communication graph wherein edges determine between which nodes migration can occur.

Each compute node is represented by a vertex in the graph, and neighbors are selected based on inter-node communication. The desired number of neighbors, or vertex degree, is a tunable parameter chosen based on the application context, as we will see later. Assume $K$ is the desired vertex degree; neighbors are then selected via the following iterative process.
\begin{enumerate}
    \item Each node computes $l$, the number of neighbors it still needs to reach the desired $K$.
    \item Every node sorts neighbors in order of decreasing communication volume and sends a request to the first $l/2$. Here $l/2$ is used to prevent unnecessarily many neighbor requests passing through the system.
    \item Nodes receive neighbor requests and check:
          \begin{enumerate}
              \item how many neighbors they have already confirmed and if this count reaches $K$ 
              \item if the confirmed neighbor count plus an additional \textit{holds} count already meets $K$
          \end{enumerate}
          If either of these conditions are satisfied, the node rejects the incoming request. Otherwise, it responds with an accept and increments \textit{holds} to reserve space for the potential neighbor.
    \item On receiving an acceptance, the original requesting node must confirm that its neighbor count and holds have not exceeded $K$ in the meantime. If not, the node accepts the incoming neighbor and sends a final acknowledgment. This produces a confirmed neighbor pairing.
    \item The above process is repeated until all nodes have $K$ confirmed neighbors each, or an upper-bound number of iterations is performed.
\end{enumerate}
The above process guarantees that every node has a maximum of $K$ neighbors, assuming it communicates with at least $K$ nodes. The $K$ parameter is runtime tunable and provided by the user. \Cref{sec:sim_eval} explores the tradeoffs of different choices of $K$. In the current implementation, the neighbor graph is reconstructed during every load balancing phase, but future work should explore neighbor graph reuse, as large-scale node-to-node communication patterns are likely to persist across many load balancing iterations.

\subsection{Virtual Load Balancing}
The second phase of load balancing involves 'virtual' load balancing, during which only load magnitudes are exchanged. This phase uses an iterative fixed-point algorithm to compute the ideal load redistribution under the neighborhood constraints imposed in the previous phase. Nodes repeatedly exchange load magnitudes with neighbors, until they converge on a load distribution wherein the load variance in each neighborhood is below a prescribed threshold. At the end of this phase, each node has a list of neighbors and how much load it should aim to send or receive from each neighbor.

The iterative method used in this phase is a variant of a simple first-order convergence implementation of this algorithm described in previous work \cite{cybenko_dynamic_1989, hu_improved_1999}, with the important constraint that only "single-hop migrations" are allowed, meaning that the load can only traverse a distance of one from its originating processor. 
Future work could explore multi hop migrations, which add complexity to the rest of the algorithm.

\subsection{Object Selection}
Once the virtual load balancing phase has computed the ideal load transfer, individual objects must be selected for migration. Our object selection strategy prioritizes communication locality with the following constraints:
\begin{itemize}
    \item For a given neighbor $n$, objects are migrated in order according to the bytes communicated with $n$ during the previous load balancing phase.
    \item When an object $o$ is selected for migration, all objects that communicate with $o$ must update their communication patterns to include communication with the new resident processor of object $o$.
\end{itemize}
This combination of constraints preserves communication locality by migrating objects that already communicate with the existing neighbor whenever possible. The second requirement is especially important when a processor migrates more objects than initially communicated with a given neighbor.




\subsection{Heirarchical Load Balancing}
The aforementioned three phases operate at the granularity of processes, called "nodes" in the rest of the paper for simlicity. Our implementation additionally supports within-process load balancing, wherein the migration patterns previously computed are refined to balance load across all threads within a process as well. Up to and including this step, object migration occurs via proxy tokens, and only once this phase is complete are all objects actually migrated to their new destinations. The within-process load balancing is algorithmically much simpler than across process and considers solely load, not communication patterns. To allow us to study the strategies at scale without needing allocation of hundreds of node, we use one process per core in our studies.

\begin{figure*}[htbp]
    \centering

    \begin{subfigure}[t]{0.3\textwidth}
        \centering
        \includegraphics[
            width=\textwidth,
            clip,
            trim=1.4in 2.5in 1.35in 2.5in
        ]{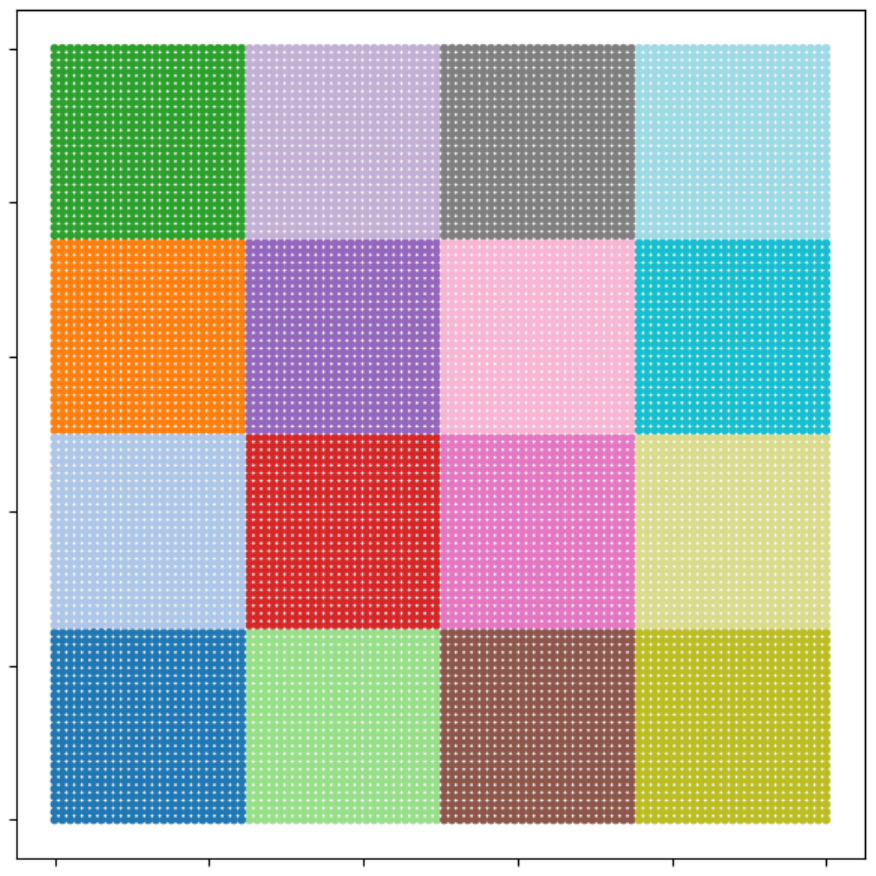}
        \caption{Original distribution: with a max to average load ratio of 1.74, external to internal communication ratio of .043.}
        \label{fig:obj-select-orig}
    \end{subfigure}\hspace{2em}
    \begin{subfigure}[t]{0.3\textwidth}
        \centering
        \includegraphics[
            width=\textwidth,
            clip,
            trim=1.4in 2.5in 1.35in 2.5in
        ]{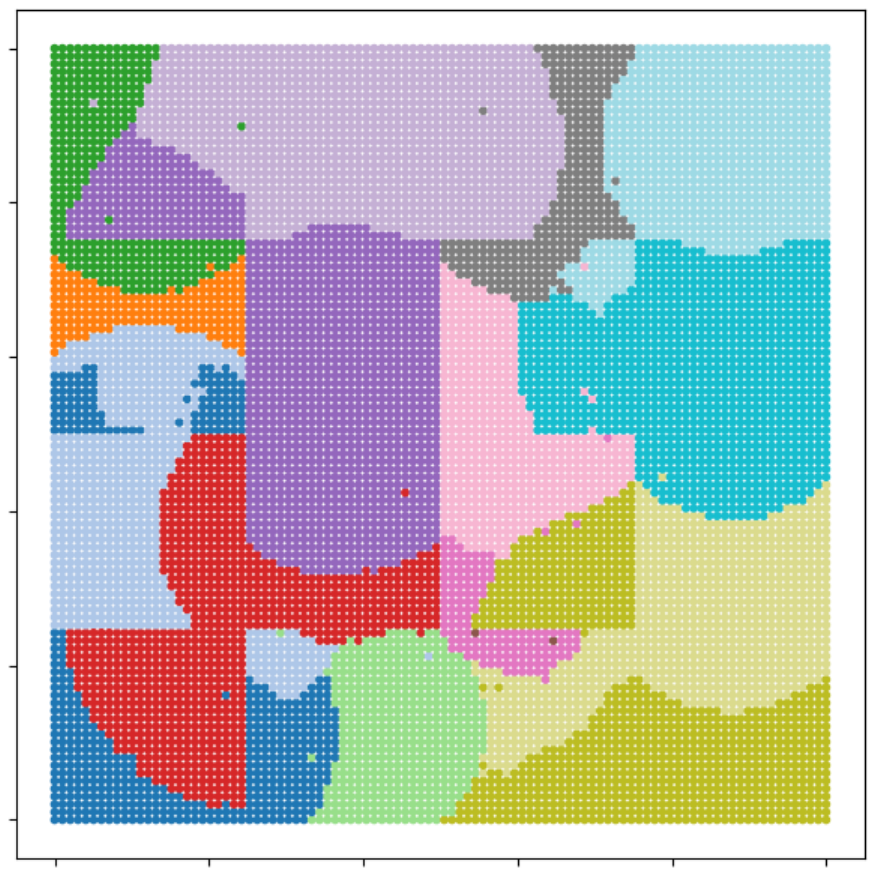}
        \caption{Coordinate-based diffusion: max to average load ratio of 1.02, external to internal communication ratio of .072.}
        \label{fig:centroid16}
    \end{subfigure}\hspace{2em}
    \begin{subfigure}[t]{0.3\textwidth}
        \centering
        \includegraphics[
            width=\textwidth,
            clip,
            trim=1.4in 2.5in 1.35in 2.5in
        ]{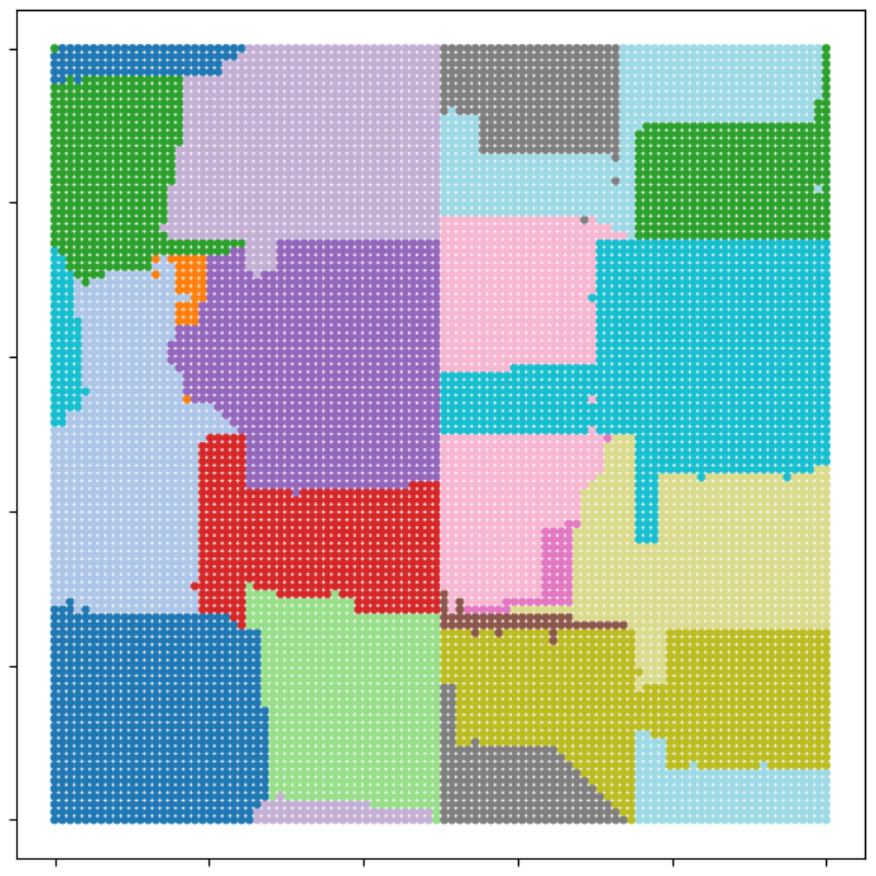}
        \caption{Communication-based diffusion:  max to average load ratio of 1.04, external to internal communication ratio of .06.}
        \label{fig:comm16}
    \end{subfigure}

    \caption{Object migration in a 2D stencil benchmark using 16 processors and an initial tiled decomposition. Imbalance is introduced synthetically so each object's load is randomly increased or decreased by 40\%. Both strategies use 4 neighbors.}
    \label{fig:16pes}
\end{figure*}

\section{Coordinate Variant}
\label{sec:coordinate}
In cases where the communication graph is not readily available, we propose an approximation to communication patterns based on object coordinates. This approximation requires that the application developer can map objects to logical positions such that inverse distance correlates with communication. This is feasible particularly in applications with a physical representation and communication locality, such as particle simulations.

The 2D stencil application is a useful example in understanding the coordinate variant, precisely because communication patterns are so closely related to chare coordinates. At the application level, we assign each chare 2D coordinates corresponding to the position of its portion of the mesh, and then the algorithm assumes that chares in proximity with each other also communicate.

To effectively use these coordinates in the absence of the communication graph, the following modifications are made to the communication-aware diffusion algorithm:
\begin{itemize}
    \item During initialization, each processor computes its \textbf{centroid}, or the average position of all of its chares.
    \item Centroids are distributed and used during neighbor selection such that all processors sort each other by inverse centroid distance. Note that this is much less scalable than the communication-based approach, because nodes must sort \textit{all} processors, instead of just those they communicate with. This limitation could be addressed in future work and is discussed in Section \ref{sec:future}.
    \item Object selection uses object distance to neighboring processor centroids, instead of communication patterns. Similarly to the communication-based variant, processor centroids are updated as objects migrate.
\end{itemize}
Importantly, the coordinate-aware variant strategy is only an approximation. The impact of these approximations on performance is discussed in evaluation.

\section{Evaluation in Simulation}
\label{sec:sim_eval}

To evaluate communication and coordinate-aware diffusion strategies in a controlled, reproducible, and efficient setting, we implement a load balancing simulation infrastructure using Charm++. The infrastructure requires as input a description of object loads, coordinates, and communication edges, which is easily generated for any Charm++ application at load balancing steps. Given this information, the simulator models various load balancing strategies independent of the application and simulates any number of processes without requiring true at-scale execution. The simulator leverages Charm++’s object-based decomposition to simulate large scale load balancing problems on a single process, wherein each chare is a stand-in for a process, enabling exploration of different imbalance patterns and migration decisions. We also use the simulator to produce the visualizations seen in \Cref{fig:intro-viz} and \Cref{fig:16pes} to provide useful intuition.

\subsection{Coordinate vs Communication-Based Diffusion}
\Cref{fig:16pes} compares our communication and coordinate-based diffusion algorithms visually in a 2D stencil application with synthetic load imbalance, over a single iteration of load balancing. Typically, coordinate-based rebalancing creates more rounded borders, due to the migration metric being derived from the distance to a processor's centroid. Communication-based rebalancing preserves original processor domain shapes better, leading to sharper linear edges.

In our current implementation, only the communication-based approach captures the periodic boundaries of the stencil application. While integrating periodic boundary support into the coordinate approach could be explored in future work, for now this scheme is limited more by the boundaries of the simulation.
This produces less optimal neighbor selection and subsequent migration, like the division of the dark blue bottom-left processor in Figure \ref{fig:centroid16}. When this dark blue processor searches for its four closest neighbors, it selects the orange processor (on the left boundary of the simulation), instead of a true communication neighbor across the periodic boundary.

Because of occasional approximations like this involved in the coordinate scheme, this approach doesn't preserve communication locality as well as the communication approach. To quantify communication locality in general, we consider the \textit{external} vs \textit{internal communication} metric, which measures how many bytes of communication move across processors vs within processor. In the example in Figure \ref{fig:16pes}, the coordinate approach produces a ratio of 0.072, with more external communication than the original Diffusion approach, achieving 0.06.

\subsection{Neighbor Count Selection}
To use the diffusion algorithm effectively, the user must select an appropriate neighbor count $K$. Intuitively, a low neighbor count restricts migration options, because nodes can only exchange load with their immediate neighbors, which can limit load balancing quality. A high neighbor count increases migration options but can be more costly to compute and additionally disrupt communication locality, if neighbors are too distant.

To visualize this tradeoff, consider a synthetic stencil application instance where processors form a 1D ring, and a single processor is heavily overloaded by a factor of 10. The initial stencil distribution 
has a load imbalance with a max to average load ratio of approximately five. Running diffusion load balancing with varying neighbor counts, \Cref{tab:neighbor-count} shows the resulting load distributions. With only 1 neighbor, the overloaded processor is unable to sufficiently diffuse its load, and as the neighbor count increases load balance improves. However, increasing neighbor count also increases the external to internal communication ratio, as more neighbors allow for more distant migrations. For example, a node may choose to migrate objects to a neighbor with which it has no communication in an attempt to distribute load. 

\begin{table}
    \centering
    \begin{tabular}{|c || c | c | c | c |}
        \hline
        Neighbor Count              & 1    & 2    & 4    & 8    \\[1ex]
        \hline
        max/avg load                & 4.9  & 1.7  & 1.3  & 1.1  \\ [1ex]
        external/internal comm (MB) & .142 & .151 & 0.25 & 0.26 \\ [1ex]
        \hline
    \end{tabular}
    \caption{Impact of neighbor count on load balancing quality in a synthetic stencil application with a single overloaded processor.}
    \label{tab:neighbor-count}
\end{table}


\subsection{Comparison with Other Strategies in Simulation}

We compare our diffusion strategies against several other load balancing algorithms implemented in our simulation infrastructure, including GreedyRefine, METIS, and ParMETIS discussed in \ref{sec:related}. GreedyRefine and METIS are already supported in Charm++ and therefore easy to plug into our simulation infrastructure. We also evaluate the MPI-based ParMETIS library, a distributed alternative to METIS that enables adaptive repartitioning, but only in simulation, for reasons explained in \Cref{sec:perf}.  


Table \ref{sim-comp} summarizes the results of this comparison across three synthetic benchmarks of increasing scale, all based on the 2D stencil application described previously. GreedyRefine consistently produces the best max to average ratio but has the worst impact on communication locality. METIS achieves good communication locality, but incurs an extremely large number of migrations. The communication-aware diffusion strategies aim to find a middle ground between these extremes.

Evaluating ParMETIS in this context proved challenging due to its parameterization, particularly tuning how much migration occurs vs how much the original partitioning should be preserved. For these experiments, we selected parameter values that produced migration counts in the same range as our diffusion strategy, though this level of parameter exploration would not be practical in general application scenarios. The complexity of these tradeoffs is particularly evident in the largest benchmark, where a better load balance could be achieved at the cost of substantially higher migration counts. Although ParMETIS serves as one of several comparative baselines, these observations highlight the practical challenges of adaptive graph partitioning strategies in dynamic load balancing scenarios.


\begin{table*}[t!] 
    \centering
    \begin{tabular}{|c || c | c c c c c ||}
        \hline
        Metric                      & Initial & GreedyRefine & METIS  & ParMETIS & Diff-Comm & Diff-Coord \\ [0.5ex]
        \hline\hline
        Benchmark 1: 8 PEs                                                                                 \\
        \hline
        max/avg load                & 1.32    & 1.00          & 1.00   & 1.01     & 1.06      & 1.05       \\
        external/internal comm (MB) & .5      & .887          & .343   & .637     & .58       & .648       \\
        \% migrations               & -       & 19.9\%        & 87.1\% & 14.6\%   & 18.9\%    & 17\%       \\ [1ex]
        \hline
        \hline\hline
        Benchmark 2: 32 PEs                                                                                \\
        \hline
        max/avg load                & 1.37    & 1.00          & 1.00   & 1.04     & 1.02      & 1.02       \\
        external/internal comm (MB) & .143    & .422          & .158   & .194     & .173      & .227       \\
        \% migrations               & -       & 18.7\%        & 98.9\% & 6.6\%    & 15.4\%    & 17.6\%     \\ [1ex]
        \hline
        \hline\hline
        Benchmark 3: 128 PEs                                                                               \\
        \hline
        max/avg load                & 1.37    & 1.00          & 1.00   & 1.20     & 1.09      & 1.14       \\
        external/internal comm (MB) & .6      & .999          & .217   & .602     & .625      & .616       \\
        \% migrations               & -       & 18.8\%        & 99\%   & 10.8\%   & 18.2\%    & 17.8\%     \\ [1ex]
        \hline
    \end{tabular}
    \caption{Comparison of load balance metrics across 5 algorithms and synthetic benchmarks with a 3D stencil communication patterns. All benchmarks use the same synthetic load imbalance injection, wherein every 1st and 2nd PEs mod 7 is overloaded, and every 3rd mod 7 is underloaded.}
    \label{sim-comp}
\end{table*}


\section{Performance}
\label{sec:perf}
To evaluate the quality of communication-aware diffusion load balancing, we use a Charm++ implementation of the PIC-inspired benchmark presented in the Parallel Research Kernels (PRK)\cite{parallelresearchkernels, prk-pic}.

This PIC PRK benchmark was designed to assess the load-balancing capabilities of parallel runtimes. It simulates a 2D mesh with periodic boundaries and fixed electromagnetic charges at each grid point. Particles are distributed throughout the mesh via a variety of different initialization modes, and at each time step particles interact with the nearby grid charges, compute the Coulomb forces exerted on the particle, and update their state accordingly. The simulation is constructed such that the horizontal displacement of any particle after one time step is determinately $(2k + 1)$ grid cells, where $k$ is a user-defined parameter. This property ensures that load imbalance patterns evolve in a predictable fashion over time, making it a useful benchmark for load balancing evaluation. PIC PRK is not a full PIC simulation, but focuses on the algorithmic components most relevant to load balancing.

We implement a Charm++ version of the PIC PRK benchmark, which decomposes the grid into cells and partitions the cells among chares in a basic 2D decomposition. It supports all relevant usage modes of the existing PIC PRK implementation (including variable grid size, particle count, and particle distribution modes), with additional parameters prescribing the number of chares used and the frequency of load balancing. We evaluate the Charm++ PIC PRK performance using no load balancing, the GreedyRefine Charm++ strategy, and our diffusion strategy. The parallel graph-partitioning alternative ParMETIS is not explored here, since its not a supported strategy in Charm++. ParMETIS is an MPI library, whereas Charm++ uses its own lower-level layers such as UCX for communication, creating an incompatibility. Using an MPI-based version of Charm++ can overcome this, but the communication cost will increase unreasonably. 

\subsection{PIC PRK Load Imbalance Patterns}
The PIC PRK benchmark supports multiple tunable parameters that impact load imbalance patterns, including the skewedness of the initial particle distribution, the horizontal speed of particles, and the vertical speed of particles. In this section, we explore our choices for these parameters to better understand the behavior of load imbalance in the application and the expected performance of load balancing strategies.

\textbf{Initial Particle Distribution:} Various initial particle distributions are supported by PIC PRK, as detailed in \cite{prk-pic}. We use the 'GEOMETRIC' distribution mode, which places particles on the grid according to an exponential distribution. In a $c$ by $c$ grid, the number of particles in the $i$-th column of grid cells is given by $A \rho^i$, 
where $A$ is a normalization constant and $\rho$ is a user-defined skew parameter. Within each column, particles are placed into rows uniformly at random. Higher values of $\rho$ lead to more skewed distributions, with more particles concentrated in the leftmost columns. Given an understanding of the initial distribution, and the aforementioned property of $k$ dependent horizontal displacement, we can fully characterize the load imbalance patterns that will arise over time.

\textbf{Particle Speeds:} The horizontal speed parameter $k$ determines how far particles move horizontally in each time step, specifically $(2k + 1)$ grid cells. Higher values of $k$ lead to more rapid movement across the grid, and $k$ must be considered carefully when determining the frequency of load balancing. We set vertical speed to 1 grid cell per timestep, and we ignore load imbalance in the vertical dimension due to the uniform distribution of particles along this axis.

For our evaluation, we use $100,000$ particles across a $1000$ by $1000$ grid over $100$ time steps, using $k=2$ and $\rho=.9$ and a 12x12 grid of chares.
Each chare spans a subgrid of approximately 80x80 grid cells.
Particles are initially exponentially concentrated on the left side of the grid, and over time they move to the right at a rate of 5 grid cells per time step, migrating from one chare to the next at regular intervals.

\textbf{Processor Decomposition:} The Charm++ PIC PRK benchmark supports two modes of initial chare-to-processor mappings: (1) a striped mapping, in which chares are assigned to processors in column-major order, and (2) a quad mapping, in which chares are assigned in contiguous 2D tiles. The striped mapping leads to more inter-processor communication, as particles frequently move between processors, while the quad mapping preserves communication locality better. Here we consider the striped mapping to better understand the column-wise load imbalance patterns of the application.

\begin{figure}
    \centering
    \includegraphics[
        width=0.64\columnwidth,
        clip,
        trim=.5in 0in .7in 0in,
        angle=270
    ]{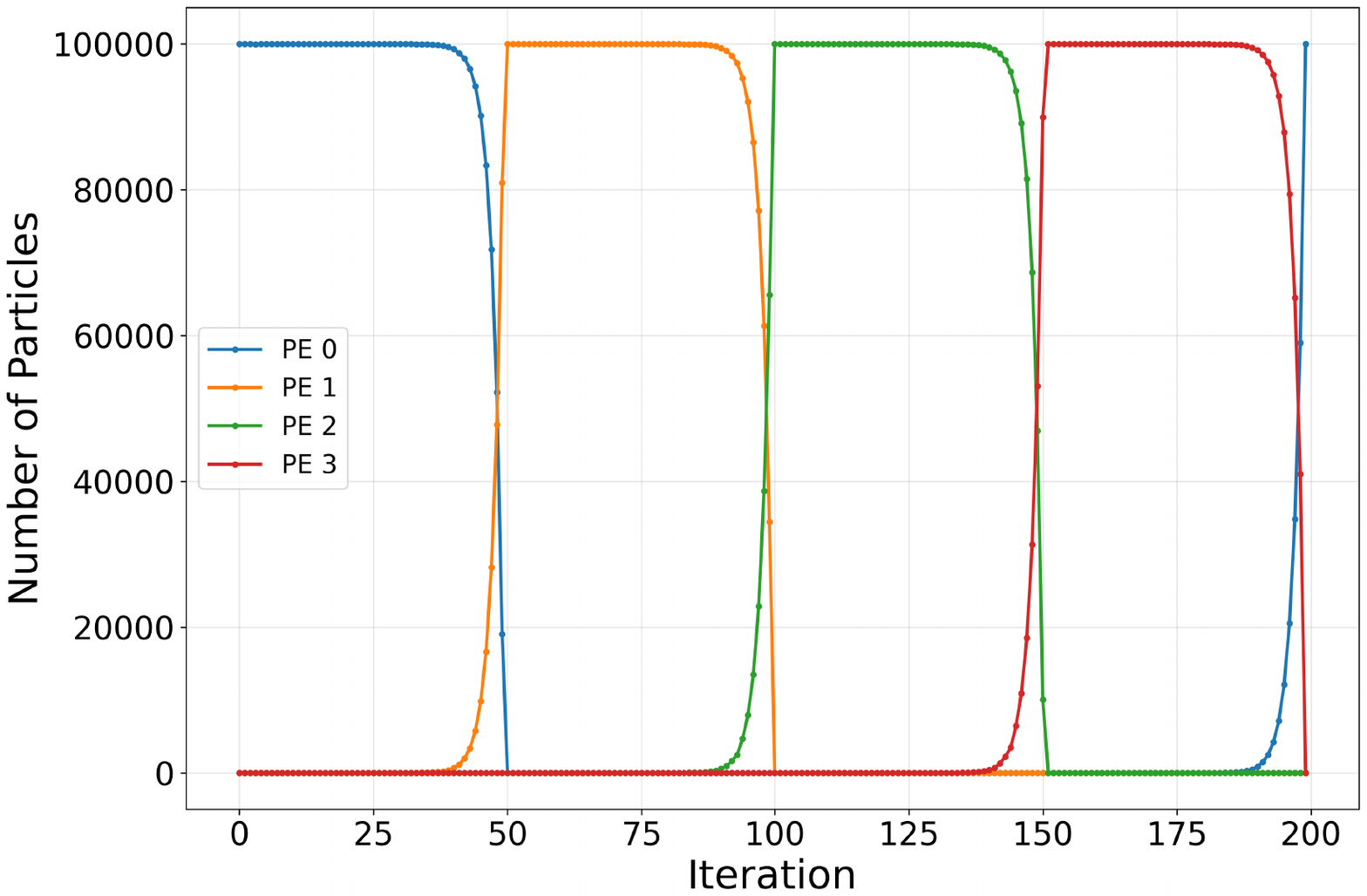}
    \caption{Evolution of particle distribution over processors over time in the PIC PRK benchmark, using $k=2$ and $\rho=.9$.}
    \label{fig:pic-nolb-particles}
\end{figure}

Figure \ref{fig:pic-nolb-particles} shows the evolution of particle counts per processor over time with 4 processors over 200 iterations to see particle migrations across the whole grid. The load imbalance pattern is clear, with processors becoming overloaded as particles move into their columns, then becoming under-loaded as particles move out.

\subsection{Particle Count Under Load Balancing}
\begin{figure}
    \centering
    \includegraphics[
        width=0.6\columnwidth,
        clip,
        trim=.5in 0in .7in 0in,
        angle=270
    ]{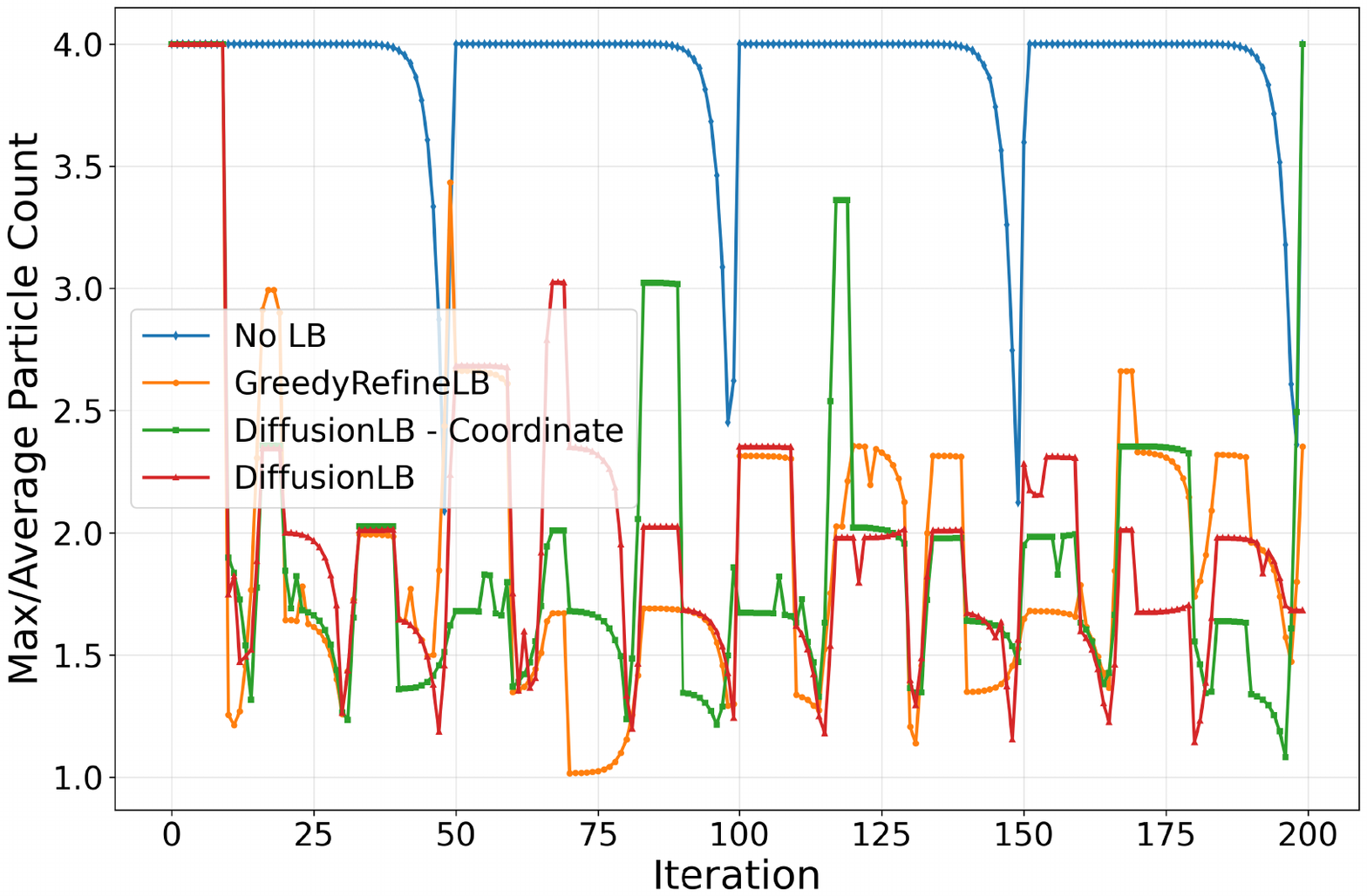}
    \caption{Ratio of max to average number of particles per processor over time in the PIC PRK benchmark, using $k=2$ and $\rho=.9$ and 4 processors. Load balancing is performed every 10 iterations and Diffusion strategies use 4 neighbors}
    \label{fig:pic-std}
\end{figure}
While particle count is only an approximation for load in the PIC PRK benchmark, and not the metric used by our load balancers, it is still useful to understand how load balancing strategies impact the distribution of particles over time. Figure \ref{fig:pic-std} shows the max to average ratio of particles per processor over time, comparing no load balancing, GreedyRefine, and Diffusion load balancing strategies. All load balancing strategies are able to maintain a much lower max to average ratio of particles per processor compared to no load balancing - in this particular use case, GreedyRefine and Coordinate-based Diffusion produce a 50\% improvement and Communication-based Diffusion produces a 48\% improvement on average.

\subsection{Multi-node Performance}

\begin{figure}
    \centering

    \begin{subfigure}[t]{0.4\textwidth}
        \centering
        \includegraphics[width=\textwidth]{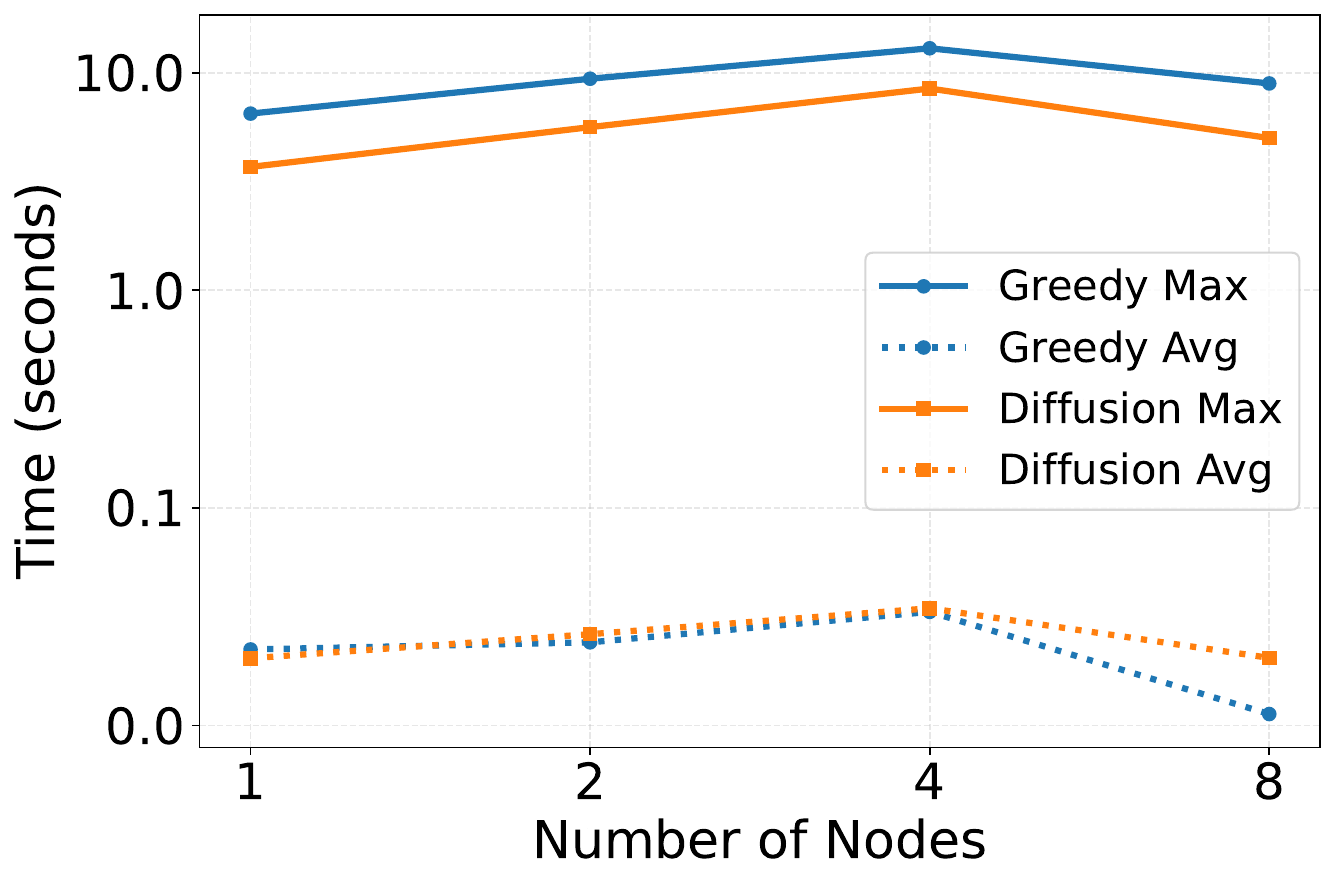}
        \caption{Communication Time}
        \label{fig:strong-scaling-1}
    \end{subfigure}\hfill
    \begin{subfigure}[t]{0.38\textwidth}
        \centering
        \includegraphics[width=\textwidth]{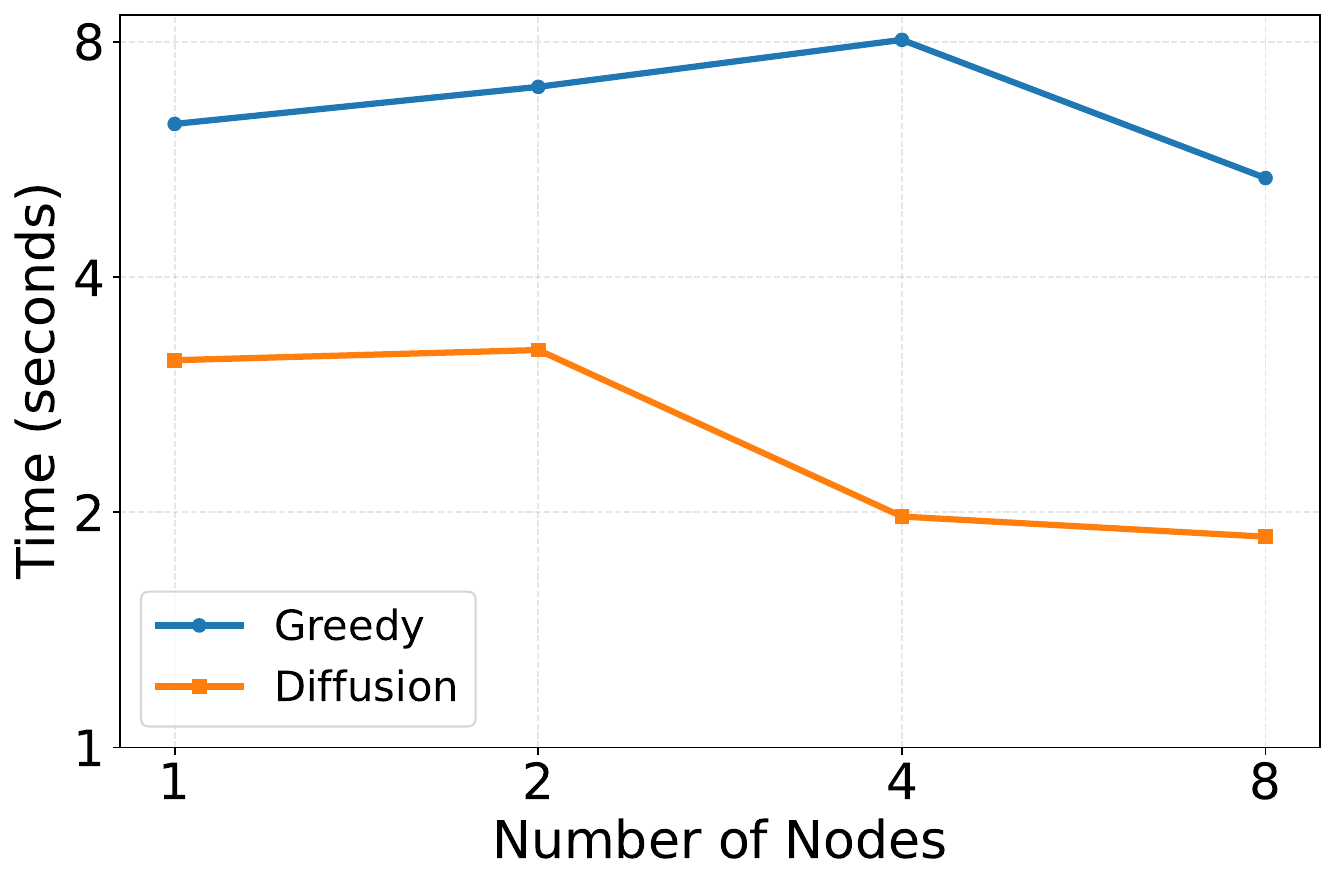}
        \caption{Load Balancing Time}
        \label{fig:strong-scaling-3}
    \end{subfigure}
    \begin{subfigure}[t]{0.4\textwidth}
        \centering
        \includegraphics[width=\textwidth]{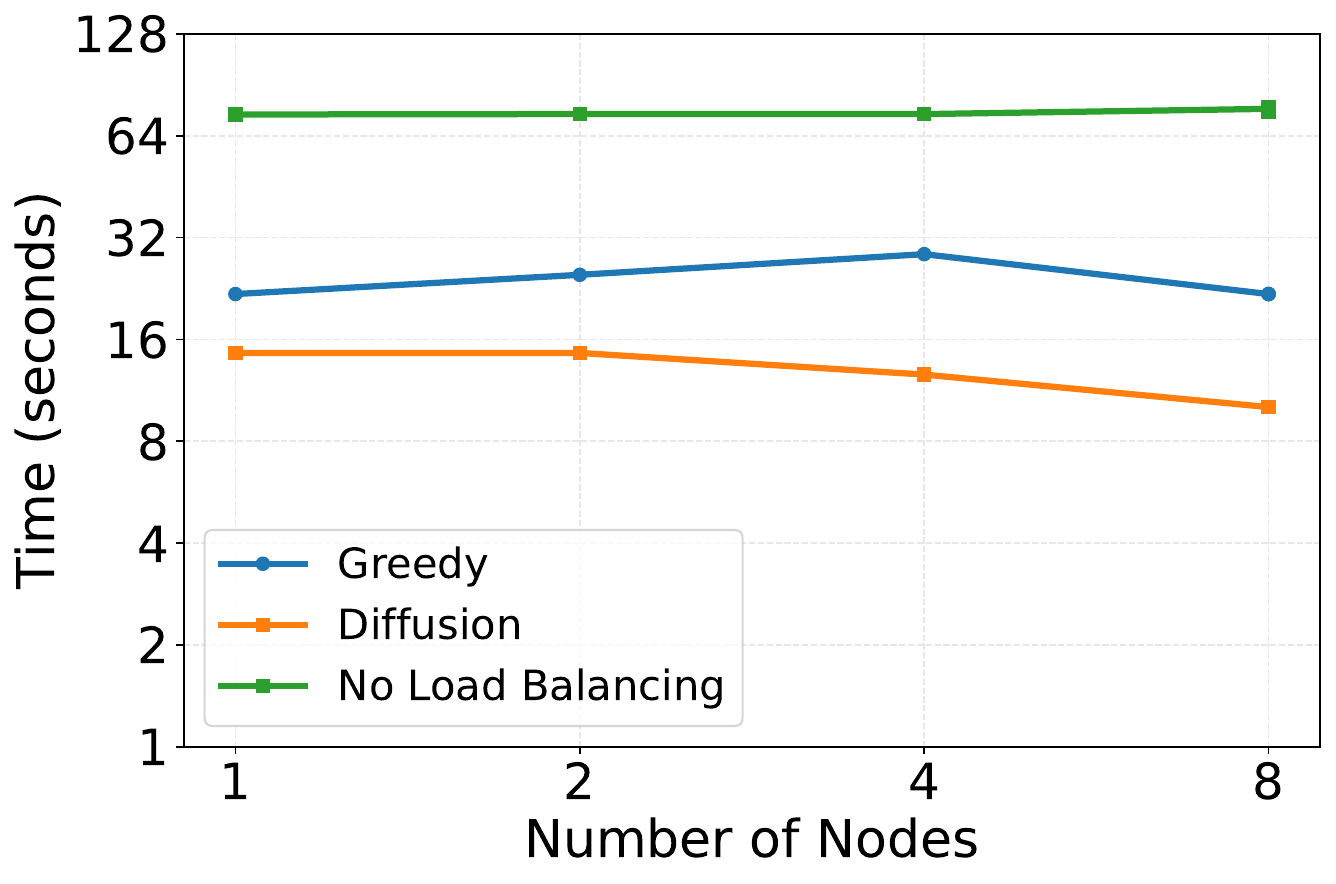}
        \caption{Overall Runtime}
        \label{fig:strong-scaling-4}
    \end{subfigure}

    \caption{Strong scaling of the PIC PRK benchmark on Perlmutter, comparing communication-based Diffusion and GreedyRefine load balancing strategies, using 10 million particles on a 6000x6000 grid with parameters $k=4$ and $\rho=.9$ and scaling the number of chares with the number of nodes.}
    \label{fig:strong-scaling}
\end{figure}

We evaluate the performance of the PIC PRK benchmark under Diffusion and GreedyRefine load balancing on up to 8 nodes of the Perlmutter supercomputer at NERSC, using 16 processes per node and 8 physical cores per process. Figure \ref{fig:strong-scaling} shows the strong scaling results of the PIC PRK benchmark, breaking down the time spent in communication, load balancing, and overall runtime. The PIC PRK benchmark exhibits such extreme load imbalance that without load balancing the application doesn't scale at all, as seen in Figure \ref{fig:strong-scaling-4}. Diffusion is able to restore scaling slightly, but is still limited by the imbalance and the communication overhead of the application. Diffusion outperforms GreedyRefine at all scales, with the performance gap widening at larger scales. At 8 nodes, Diffusion achieves a 2x speedup over GreedyRefine and a 7x speedup over the original runtime.

\begin{figure}[htbp]
    \centering
    \begin{subfigure}[t]{0.45\textwidth}
        \centering
        \includegraphics[
            width=\textwidth,
        ]{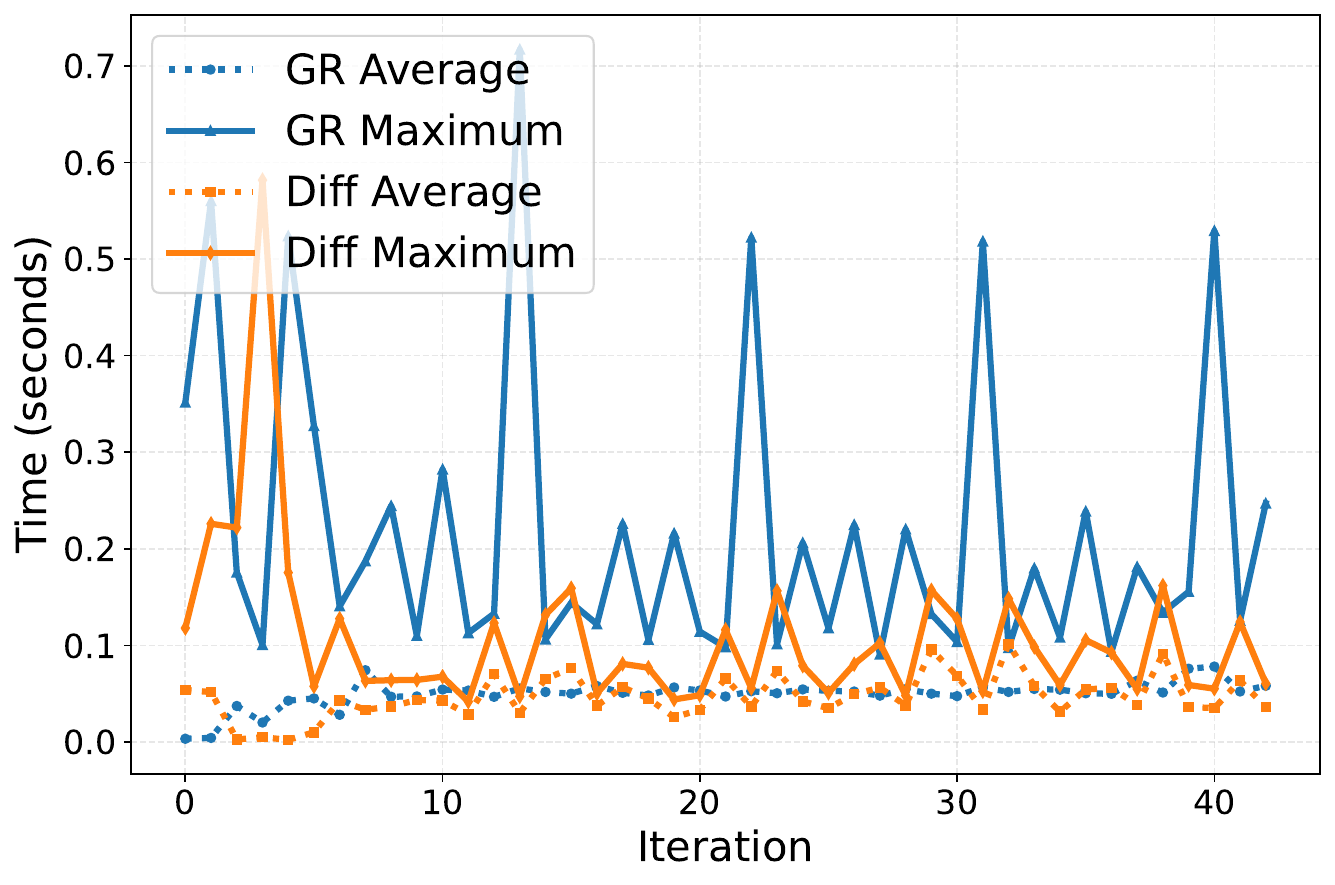}
        \caption{Communication Time}
        \label{fig:8nodecomm}
    \end{subfigure}
    \begin{subfigure}[t]{0.45\textwidth}
        \centering
        \includegraphics[
            width=\textwidth,
        ]{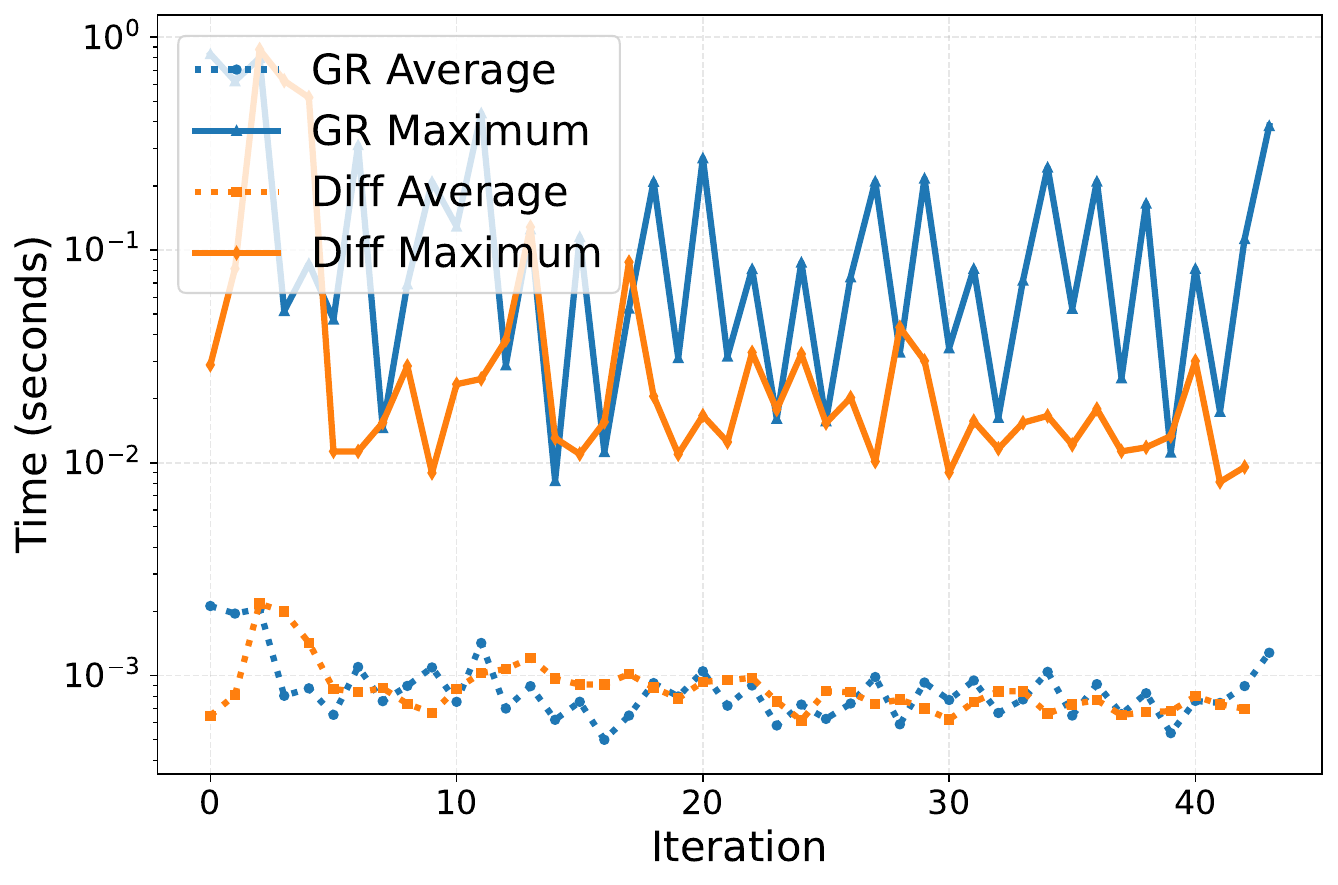}
        \caption{Computation Time}
        \label{fig:8nodecomp}
    \end{subfigure}

    \caption{Communication and computation time per process over 100 load balancing phases in the PIC PRK benchmark on 8 nodes of Perlmutter, using 10 million particles on a 6000x6000 grid with parameters $k=4$ and $\rho=.9$ and 200x100 chares. Load balancing is performed every 5 iterations to compare communication-based Diffusion and GreedyRefine.}
    \label{fig:8node-commcomp}
\end{figure}

To evaluate the impact of communication locality, we break down the 8 node results further. Figure \ref{fig:8node-commcomp} shows the time spent in communicaton and computation averaged across PEs for 100 iterations, with load balancing performed every 5 iterations. While GreedyRefine manages to bring down the maximum communication time occassionally, 
it exhibits spikes and a much higher communication cost compared to Diffusion. On average, Diffusion achieves a 2x speedup in maximum communication time per PE compared to GreedyRefine.

The computation time on average is about the same under both algorithms, as expected. 
However, the max computation time is improved with Diffusion, achieving a 2.5x speedup over GreedyRefine on average, as Diffusion is better able to adapt with the simulation dynamics and produce more consistent load balance across iterations.

\section{Future Work}
\label{sec:future}
Most notably, future work should explore the performance of Communication-Aware Diffusion load balancing in full-scale applications beyond the PIC PRK benchmark. Applications with complex and dynamic communication patterns would provide a more rigorous test of the proposed strategy. Specifically ChaNGa \cite{changa}, a Charm++ cosmological simulation, would benefit from improved load balancing strategies that preserve communication locality. Future work will evaluate Diffusion against ChaNGa's custom load balancing strategy, with a focus on the preservation of communication locality through many iterations of load balancing.

Additionally, it would be valuable to compare our strategies against a wider array of Charm++ strategies, in addition to GreedyRefine explored here. Specifically, ParMETIS repartitioning showed promise in our simulation studies, but is not currently available in Charm++, due to it being an incompatible MPI-based library. Integrating it in Charm++ would require a larger engineering effort and is left for future work.

The current coordinate-based variant requires each processor to sort all other processors by centroid distance during neighbor selection, which limits scalability as opposed to the communication-based variant. Future work could explore more efficient spatial indexing techniques such as space-filling curves (SFC) or octree-based k-nearest neighbor algorithms to identify neighboring processors more efficiently.

While our proposed algortihm is in theory applicable to many task-based contexts, it is difficult to port and evaluate load balancing strategies across different runtimes due to their tight integration with runtime-specific features and APIs. Future work will explore a general framework that abstracts the core diffusion strategy from runtime-specific details, enabling integration with other task-based runtimes and facilitating broader adoption and comparative studies.

\section{Conclusion}
\label{sec:conclusion}

We present a communication-aware diffusion load balancing algorithm designed for irregular, communication-intensive parallel applications in over-decomposed runtime systems. Additionally, we present a coordinate-based variant for scenarios where explicit communication patterns are unavailable, using spatial proximity as a proxy for communication relationships, and comparing the outcomes.

Through simulation-based evaluation, we demonstrate that our diffusion strategies achieve competitive load balance compared to established methods like GreedyRefine and graph partitioning approaches, while maintaining significantly better communication locality and requiring fewer object migrations. Performance evaluation on the PIC PRK benchmark shows that communication-aware diffusion achieves overall speedup compared to GreedyRefine and lower average communication time per chare, demonstrating the practical benefits of preserving locality in dynamic load balancing.

Our results indicate that incorporating communication awareness into a diffusion-style algorithm can improve load balance while limiting communication overhead in this setting. Although our evaluation is limited in scale, these results suggest that diffusion-based approaches are a promising load balancing direction for communication-intensive applications.

\bibliographystyle{IEEEtran}
\bibliography{diffusion}
\end{document}